\documentclass[twocolumn, twocolappendix]{aastex701}
\usepackage{threeparttable}
\usepackage{amsmath}

\begin{document}

\title{Across the Neptune Landscape: Distinct Core Mass, Planet Mass, and Envelope Mass Fractions for Neptunes in the Desert, Ridge, and Savanna}

\author[0009-0002-2836-3447]{Sharone E. Abhilash}
\affiliation{Department of Chemical \& Physical Sciences, University of Toronto Mississauga\\ 3359 Mississauga Road, L5L 1C6 Mississauga ON, Canada}
\email{sharone.abhilash@mail.utoronto.ca}

\author[0000-0002-6076-5967]{Marta L. Bryan}
\affiliation{Department of Astronomy and Astrophysics, The Pennsylvania State University, 525 Davey Lab, University Park, PA 16802}
\affiliation{Center for Exoplanets and Habitable Worlds, 525 Davey Laboratory, The Pennsylvania State University, University Park, PA 16802}
\email{martalbryan@psu.edu}

\author[0000-0002-9849-5886]{J. J. Zanazzi}
\affiliation{Department of Astronomy and Astrophysics, The Pennsylvania State University, 525 Davey Lab, University Park, PA 16802}
\affiliation{Center for Exoplanets and Habitable Worlds, 525 Davey Laboratory, The Pennsylvania State University, University Park, PA 16802}
\email{jxz224@psu.edu}

\begin{abstract}
Neptune-mass planets at sub-three-day orbital periods are strikingly rare (``Neptune Desert''), but just exterior to three-day periods lies an occurrence-rate over-density (``Neptune Ridge'') that becomes more sparse beyond six-day periods (``Neptune Savanna''). Whether the compositions and dynamical histories of these three Neptune-mass populations are similar or distinct remains unclear. This paper combines MESA planetary evolution models with Bayesian statistics to calculate the core masses and envelope mass fractions (EMFs) of 215 Neptunes with measured masses and radii, inferring population-level demographic properties of Neptunes in the Desert, Ridge, and Savanna. We show that each population displays its own exemplary attribute: Desert Neptunes have more massive cores, Ridge Neptunes have larger EMFs, and Savanna Neptunes are generally less-massive overall. After re-affirming that Neptune Desert and Ridge hosts have systematically super-solar spectroscopic Gaia DR3 metallicities while Savanna host values peak at solar, we find Savanna Neptunes orbiting metal-poor hosts seem to be ``scaled-down'' versions of their metal-rich counterparts, with lower core masses, planet masses, and EMFs. We posit three distinct origins for the Neptune populations: 1) Savanna Neptunes formed gas-poor \textit{in-situ}; 2) Ridge Neptunes formed gas-rich at wide separations, delivered to short periods following high-eccentricity migration; and 3) Desert Neptunes are likely barren cores of former hot Jupiters, placed in the Desert either via high-eccentricity migration followed by dynamical tides or Roche-lobe overflow. Future Rossiter-McLaughlin measurements can distinguish between these Desert placement scenarios.

\end{abstract}

\section{Introduction} 
\label{sec:Intro}
 
The demographics of exoplanets in the mass--period parameter space reflect how those planets form, migrate, and evolve. One intriguing feature in this space is a prominent under-density of close-in Neptune-mass planets ($\sim$10--100 M$_{\oplus}$) at orbital periods shorter than $\sim$3.2 days, known as the ``Neptune Desert'' \citep{Szabo_2011,Beauge_2013,Mazeh_2016}. This scarcity is not an observational bias, since current transit and radial velocity (RV) surveys are easily sensitive to planets of this mass on such short orbits. 

The upper and lower boundaries of the Neptune Desert are thought to be carved by a combination of processes. Photoevaporation can set the lower boundary, stripping the atmospheres of small planets (e.g. core masses of $\sim$10--15$\,M_\oplus$), but this mechanism cannot strip the envelopes of massive, gas-rich planets \citep{Owen_Jackson_2012,Owen_lai_2018}. The upper boundary instead is consistent with delivery to short orbital periods via high-eccentricity migration (HEM), in which a giant planet excited onto an eccentric orbit tidally circularizes at a semi-major axis of $\sim$twice the Roche radius \citep[e.g.][]{Ford_Rasio_2006, MatsakosKonigl2016, Owen_lai_2018, Zanazzi_2026}. 

More recently, new planet discoveries dominated by TESS combined with dedicated RV follow-up have yielded a small but growing population of planets in this Desert \citep[e.g.][]{Barragan2019,Burt2020,Naponiello2023,Polanski2024}. Preliminary characterization of this population indicates that these ``Desert Neptunes'' are remarkably dense \citep[e.g.][]{Castro-Gonzalez+2024_density, Zanazzi_2026}, and have stellar metallicities that are elevated in comparison to their longer-period Neptune counterparts, but consistent with hot Jupiter host metallicities \citep{Vissapragada_2025}. Taken together, these observations raise the question: are Desert Neptunes a continuation of the broader Neptune population that somehow survived some destructive process, or are they instead gas giants whose envelopes have been stripped, leaving behind Neptune-mass cores?

Here we investigate whether Desert Neptunes are a continuation of the broader Neptune population. At wider separations, Neptune-mass planets fall into one of two additional populations. Immediately beyond the Desert, at $3.2 < P < 5.7$ days lies a pile-up of Neptune-mass planets called the ``Neptune Ridge'' \citep{Castro_2024}, followed at longer periods by the more sparsely populated ``Neptune Savanna'' \citep{Bourrier_2023}. Preliminary observations of these three Neptune populations hint at trends; Desert and Ridge planets have similarly elevated host star metallicities compared to planets in the Savanna, and Ridge Neptunes additionally have a handful of non-zero eccentricity and misaligned obliquity measurements, which point to a dynamically hot history \citep{Correia_2020, Bourrier_2023, Yee_2025}.

In this work we take all planets across this Neptune landscape that have both measured masses and radii, infer the core mass, Envelope Mass Fraction (EMF), and planet mass of each Neptune, using a grid of planetary evolution models with posteriors inferred using Baysian statistics, and quantify comparisons of these distributions across the Desert, Ridge, and Savanna. Section~\ref{sec:Methods} describes our sample of Neptunes with measured masses and radii and details the evolutionary models and MCMC framework that we assemble and employ for each individual planet in our sample. Section~\ref{sec:results} presents the population-level distributions of core mass, EMF, and planet mass across the Desert, Ridge, and Savanna. Implications on the formation and evolution of Neptune-mass planets are discussed in Section \ref{sec:disc}, and summarize our main findings in Section~\ref{sec:conc}. 

\section{Methods}
\label{sec:Methods}

\subsection{Exoplanet Sample}
\label{sec:Observations}

\begin{figure}[t]
    \centering
\includegraphics[width=\columnwidth]{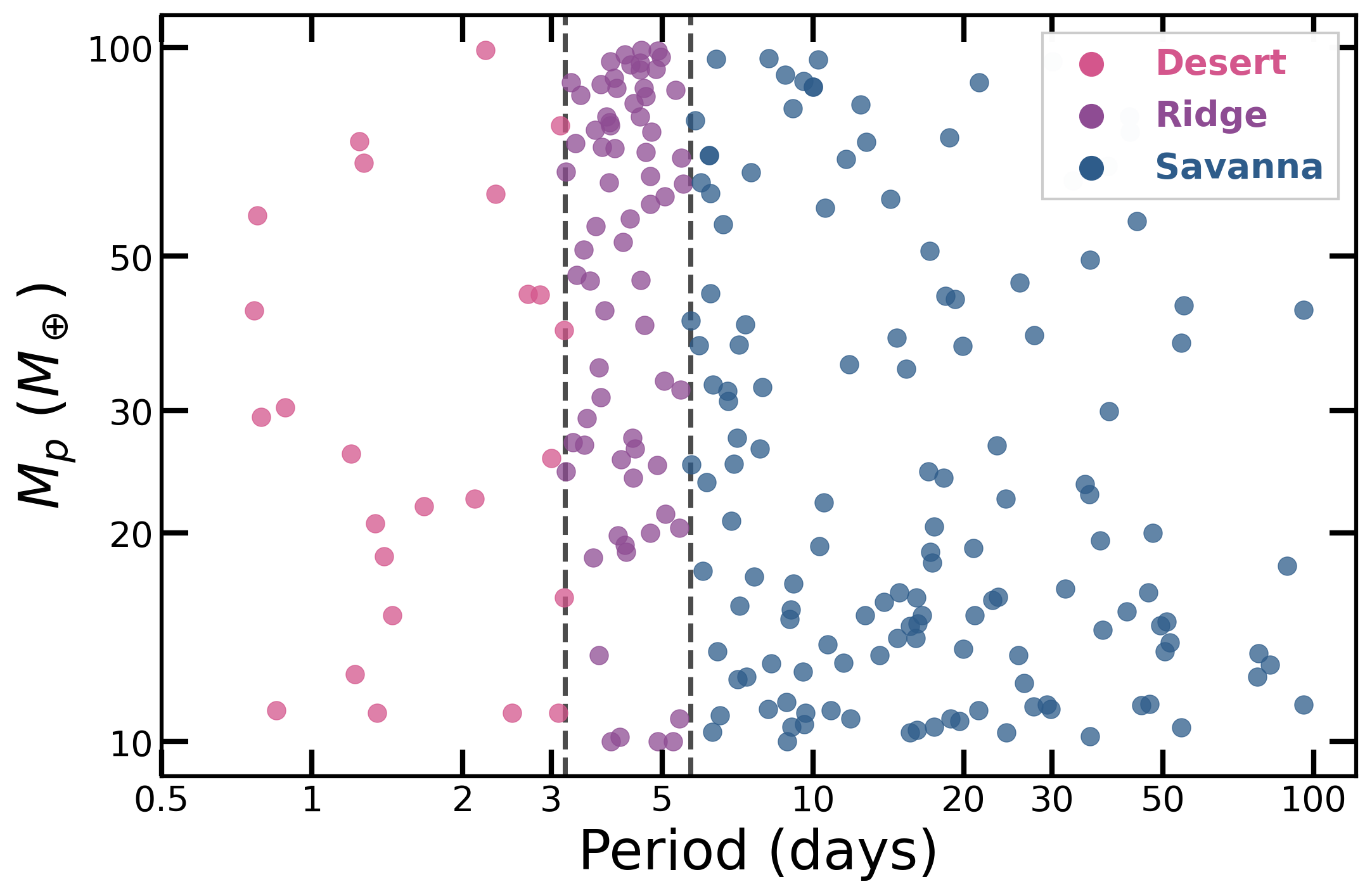}
    \caption{\textbf{Our exoplanet sample} consisting of Desert Neptunes ($P_{\rm orb}$ $<$ 3.2 days; pink), Ridge Neptunes (3.2 $\leq$ $P_{\rm orb}$ $<$ 5.7 days; purple), and Savanna Neptunes ($P_{\rm orb}$ $\ge$ 5.7 days; blue), with dashed lines denoting the 3.2 and 5.7 day boundaries \citep{Castro_2024}.
    See Section~\ref{sec:Observations} for selection criteria.}
    \label{fig: sample}
\end{figure}

Our goal in this work is to determine the core masses and EMFs of exo-Neptunes in the Neptune Desert, Ridge, and Savanna. To do this, we first construct our planet sample using the NASA Exoplanet Archive \citep{NASAArchive}, and require all planets to have measured mass and radius measurements. Masses must lie between 10--100 $M_\oplus$ to a precision better than 30\%, and the most recent value with smallest uncertainty is adopted when multiple mass measurements are available. We further weed out M-dwarf hosts, requiring the host star mass to lie above $0.6 \, M_\odot$ (see Appendix~\ref{sec:appendix_stellarmass} for discussion of stellar mass distributions for our three Neptune populations). These selection criteria yield 231 Neptune-mass planets orbiting F/G/K stars, with 25 in the Desert ($<$3.2 day, e.g. \citealt{Castro_2024, Vissapragada_2025}), 70 in the Ridge (3.2 to 5.7 day, e.g.~\citealt{Castro_2024, Vissapragada_2025}), and 136 in the Savanna ($>$5.7 day, e.g.~\citealt{Bourrier_2023}; Fig.~\ref{fig: sample})

\subsection{Planet Structure Models}
\label{sec:Models}

Placing a solar composition gaseous envelope atop a $10 \, M_\oplus$ core directly corresponds to a specific planetary size. If the planet mass consists of two components, a gaseous envelope and a solid core, a radius measurement constrains the gas-to-core mass fraction. This sub-section discusses our idealized Neptune internal structure model, consisting of a solar composition envelope lying on top of an inert, iron-silicon core. Later, we benchmark our core-mass constraints to the bulk metal masses obtained by other models and demonstrate remarkable consistency given differences in model complexity. Our inlists are publicly available on Zenodo, which we will post upon acceptance of our manuscript.

We construct planetary models with the Modules for Experiments in Stellar Astrophysics (MESA; version 24.08.1) code, and utilize the Modules for Experiments in Stellar and Planetary Astrophysics (MESPA)\footnote{\url{https://github.com/uzhplanets/mespa}} implementation of the \cite{Chabrier_2021} hydrogen-helium equation of state, with a 50--50 water--rock heavy element mixture. Gas composition is roughly solar, with hydrogen, helium, and metal mass fractions of $\{X,Y,Z\} = \{0.74, 0.24, 0.02\}$. Default MESA opacities for solar-composition gas are used, interpolating tabulated values from \cite{GrevesseSauval1998} for high temperature, and \cite{Freedman+2008} for low temperature. Inert cores bind gas to the planet, with the core radius determined using the \cite{Fortney_2007} mass-radius relation, assuming a two-thirds rock to one-third iron rock-mass-fraction (roughly Earth composition).

Our procedure for building and evolving Neptunes mirrors that of \cite{Chen_Rogers_2016} (see also e.g.~\citealt{Batygin+2013, Owen_Wu_2013, OwenWu2016}). An adiabatic, hot, Jovian initial model is first created, with a mass and radius of $1 \, M_{\rm Jup}$ and $2 \, R_{\rm Jup}$. Inert cores with differing masses are then inserted using the \texttt{relax\_core} option from MESA. Once gas adjusts to the presence of a core, excess gas is removed using MESA's \texttt{relax\_mass\_scale} to create a planet with total mass between 10--100 $M_\oplus$. To ensure Neptunes have high initial entropies, we add an artificial core luminosity of one times the planet's surface luminosity, and allow the planet reinflate over 1000 years before removing the artificial luminosity. We also experimented with the \cite{Chen_Rogers_2016} reinflation prescription of three times the surface luminosity when total masses exceeded $17 \, M_{\oplus}$, and found these two prescriptions gave planet radii that differed by less than 1\%. Irradiation from the host star is then added via the \texttt{relax\_initial\_irradiation} option, depositing incident day-side fluxes of $F_{\rm irrad}$ to fixed column depths of $\Sigma = 100 \ {\rm g} \, {\rm cm}^{-2}$, which approximates heating the upper regions of a sub-Saturn atmosphere to pressures of order $P \sim g \Sigma \sim 0.1 \ {\rm bar}$. Although irradiation travels less deep for smaller planets, varying $\Sigma$ between 1--100 ${\rm g} \, {\rm cm}^{-2}$ affects sub-Saturn radii by less than a few percent \citep[e.g.][]{Valsecchi+2015}. Our model grid consists of core masses between 6--100 $M_\oplus$, EMFs between 0.5--98 \%, and incident irradiation fluxes between $10^5 - 2 \times 10^9 \ {\rm erg} \, {\rm s}^{-1} \, {\rm cm}^{-2}$ (Table~\ref{tab:grid}). The range of core masses, EMFs, and irradiation fluxes are chosen to encompass the bulk densities and irradiation levels present in our exoplanet sample. Appendix~\ref{sec:appendix_models} includes a list of grid parameters for which MESA does not converge on a model in hydrostatic equilibrium (Tab.~\ref{tab:missing_models}).

\begin{figure}[t]
    \centering
\includegraphics[width=\columnwidth]{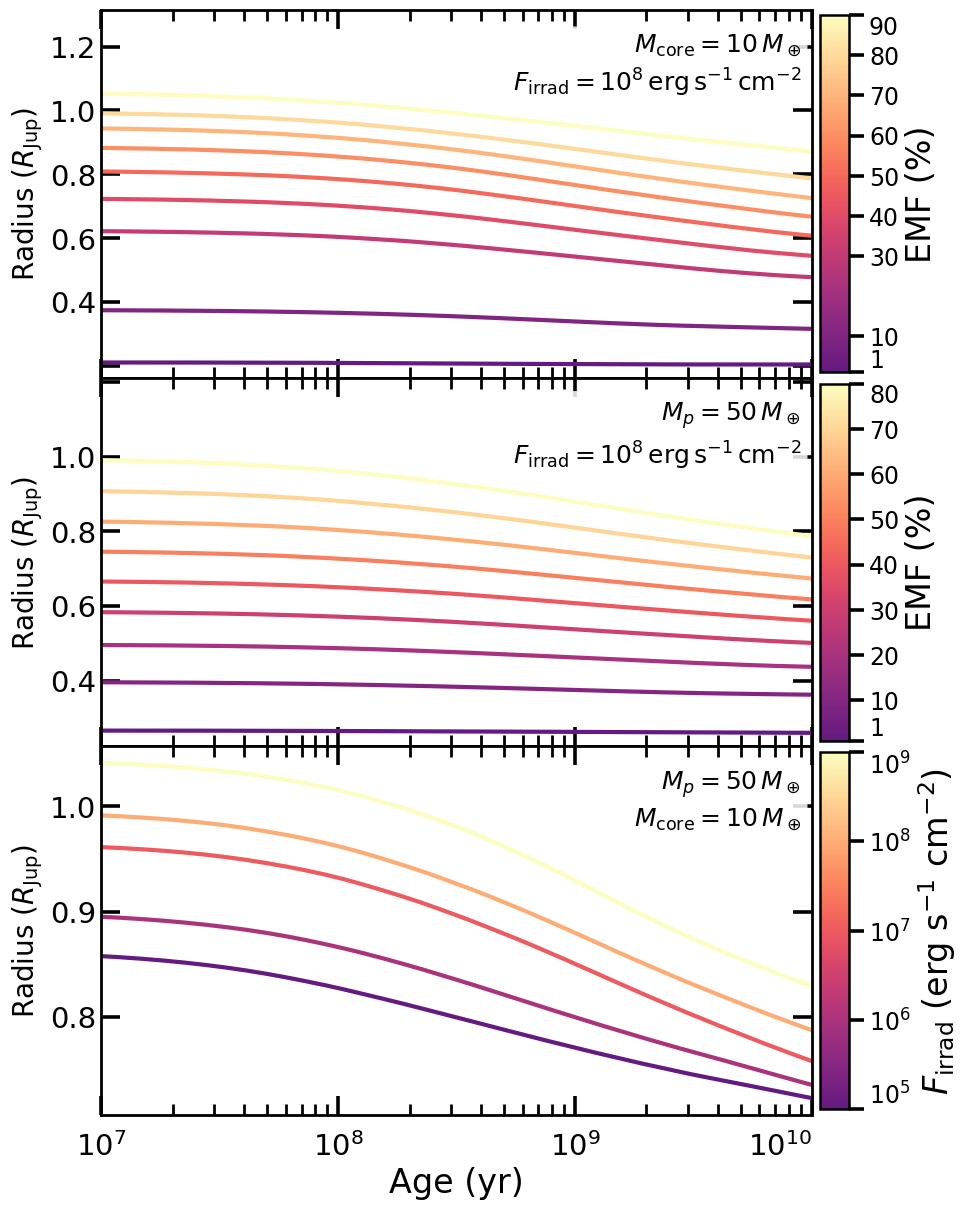}
    \caption{\textbf{EMF has the largest effect on planet radius in our models.} Radius versus age from our MESA models, varying EMF (top), core mass (middle), and irradiation (bottom). This is consistent with the models of \citet{Lopez_Fortney_2014}.}
    \label{fig: model_ev}
\end{figure}

After creating and irradiating our Neptunes, we evolve the planet for 10 Gyr, allowing its radius to contract as it cools. In addition to entropy within the envelope following formation, we inject luminosity into the base of the envelope due to the thermal inertia and radiogenic heating from a silicate-iron core \citep[e.g.][]{Nettelmann+2011, Lopez+2012}:
\begin{equation}
L_{\rm core} = -c_{\rm v,core} M_{\rm
core}\frac{dT_{\rm core}}{dt} + L_{\rm radio},
\end{equation}
We take the core specific heat $c_{\rm v,core}=1.0\times10^{7}\,\mathrm{erg\,g^{-1}\,K^{-1}}$ \citep{Valencia+2010}. $L_{\rm radio}$ is the radiogenic heating from the decay of $^{238}$U, $^{235}$U, $^{232}$Th, and $^{40}$K, with abundances, specific heat production rates, and half-lives from \citet{geo}. Each model radius is calculated at eight intermediate ages from 100 Myr to 10 Gyr (Table \ref{tab:grid}).

\begin{table*}[t]
    \centering
    \begin{threeparttable}
    \caption{Parameter values used to compute the MESA interior structure models.}
    \label{tab:grid}
    \begin{tabular}{@{\extracolsep{\fill}}ll}
        \hline
        Parameter & Values \\
        \hline
        Core mass ($M_\oplus$)
        & 6, 7, 8, 9, 10, 15, 20, 25, 30, 35, 40, 45, 50, 55, 60, 65, 70, 75, 80, 85, 90, 95, 100\\
        Envelope mass fraction (\%)
        & 0.5, 1, 5, 10, 15, 20, 25, 30, 35, 40, 50, 55, 60, 65, 70, 75, 80, 90, 95, 98\\
        Irradiation ($\mathrm{erg\,s^{-1}\,cm^{-2}}$)
        & $10^5$, $3\times10^5$, $10^6$, $3\times10^6$, $10^7$, $3\times10^7$, $10^8$, $3\times10^8$, $10^9$, $2\times10^9$\\
        Age (Gyr)
        & 0.1, 0.5, 1, 3, 5, 7, 8, 10 \\
        \hline
    \end{tabular}
    \begin{tablenotes}
        \small
        \item[] \textit{Note.} -- We compute planetary interior structure models at each combination of core mass, envelope mass fraction, irradiation, and age listed above, and interpolate between these models to infer a planet's radius as a function of these four parameters.
    \end{tablenotes}
    \end{threeparttable}
\end{table*}

The EMF has the strongest affect on Neptune radii, in agreement with the findings of \cite{Lopez_Fortney_2014}. Varying ${\rm EMF} = 1 - M_{\rm core}/M_p$ or the core mass $M_{\rm core}$ has a much larger affect on radius than cooling with age (Fig. \ref{fig: model_ev} top and middle panels). Irradiation flux $F_{\rm irrad}$ has a similarly small impact on size, with $F_{\rm irrad}$ and Kelvin-Helmholtz contraction both causing planet radii to typically vary by $<$ 20\% (Fig.~\ref{fig: model_ev} bottom panel). Our planet radii are those of our model's photosphere, and neglect the difference in radii inferred by the transit depth, typically of order a few pressure scale-heights. Given that Neptune radii are primarily determined by their core masses and EMFs, we discuss in the following subsection how we use measured masses and radii to infer these two planet properties for each Neptune in our sample.

\subsection{MCMC Composition Posteriors}

\begin{figure}[t]
    \centering
\includegraphics[width=\columnwidth]{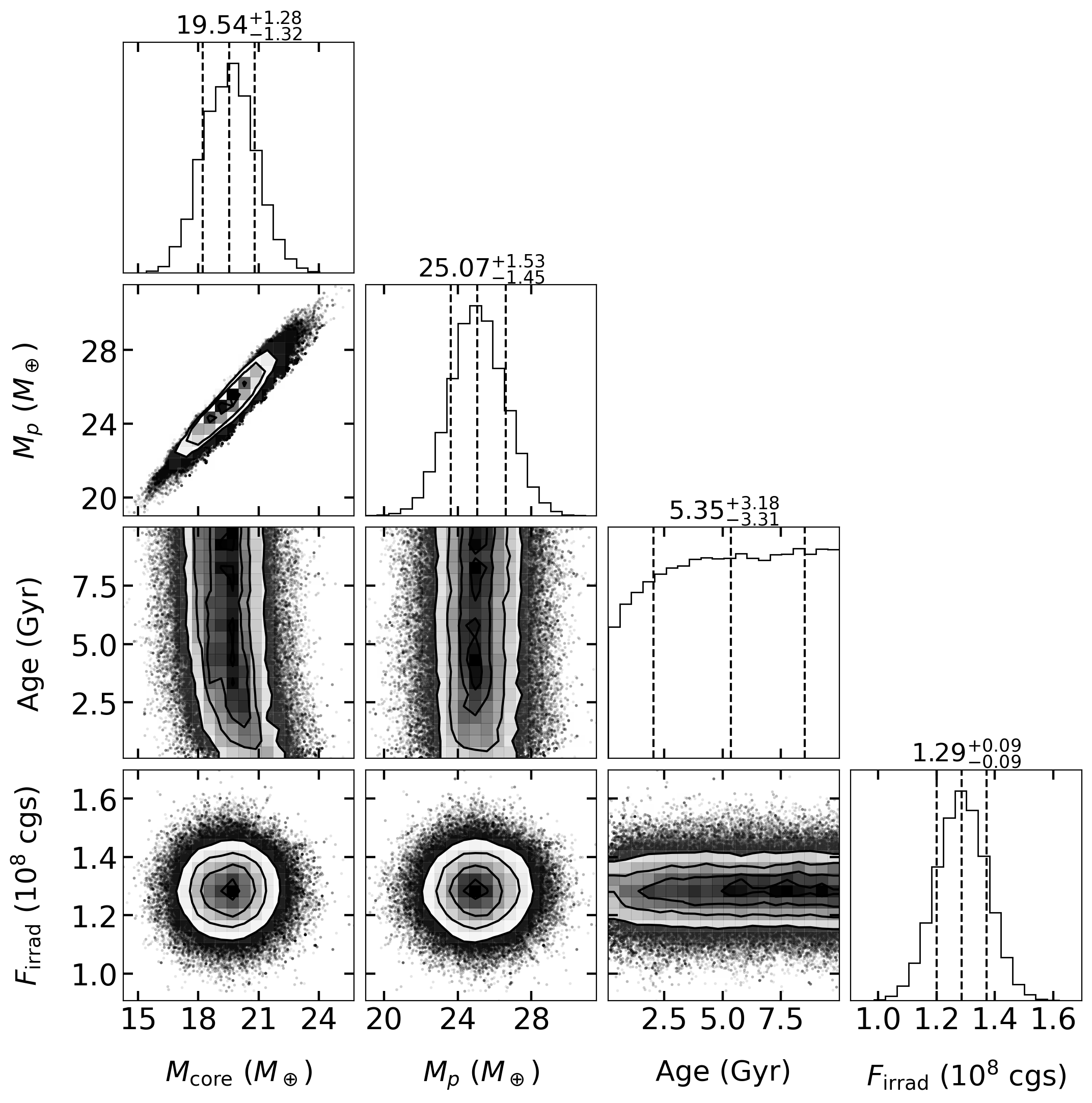}
    \caption{\textbf{Example posterior distribution from our MCMC fitting procedure.} Posterior distributions for $M_\mathrm{core}$,  $M_{p}$, age, and $F_\mathrm{irrad}$ for HAT-P-11 b, a Ridge planet in both our sample and in the planet sample of \citet{Chachan_2025}. The dashed lines mark the median and 16th/84th percentiles of each distribution. }
    \label{fig: p2 mcmc}
\end{figure}

Given an observed mass and radius for a Neptune in our sample, we can use our model grid in a Markov Chain Monte Carlo (MCMC) framework to compute posterior distributions for core mass and corresponding EMF for that planet. For model values that lie between our MESA grid points, we use \texttt{LinearNDInterpolator} within \texttt{SciPy} to perform piecewise linear interpolation over data in $\{M_{\rm core}, M_p, \log_{10} F_{\rm irrad}, {\rm Age}\}$ 4D space. Flat priors are adopted for core masses between 6--100 $M_\oplus$ and ages between 0.1--10 Gyr, restricting the ${\rm EMF} = 1 - M_{\rm core}/M_p$ to fall between 0.1--98\%. Gaussian priors are placed on the planet mass and irradiation, centered on their measured values with widths equal to their reported uncertainties (see Table \ref{tab:sample} for details). We assume a gaussian likelihood when the model radius is compared to that observed, including the radius's measurement uncertainty. Each chain is run using \texttt{emcee} with 64 walkers for 5000 steps, discarding the first 500 steps as burn-in. The median and 1$\sigma$ uncertainties from the posteriors are reported in our $M_{\rm core}$ and EMF fits (Table \ref{tab:sample}). Figure \ref{fig: p2 mcmc} displays a representative set of posteriors for the Ridge Neptune HAT-P-11 b; although the age posteriors are wide ($5 \pm 3$ Gyr), we obtain a relatively narrow core mass posterior $M_{\rm core} = 20 \pm 1 \ M_\oplus$ for this $25 \ M_\oplus$ planet.

\begin{figure}[t]
    \centering
\includegraphics[width=\columnwidth]{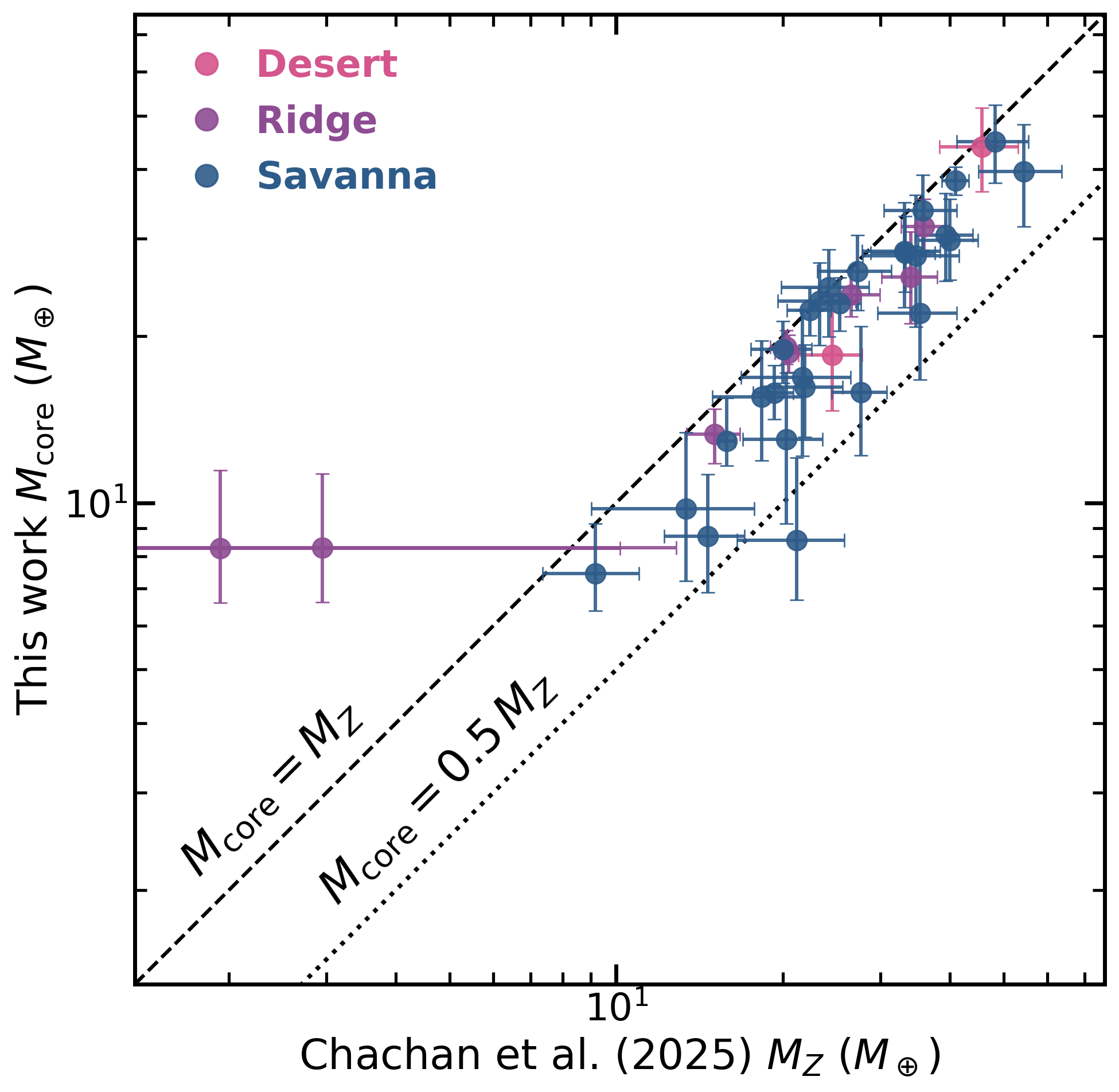}
    \caption{\textbf{Comparison of this work's core masses with the heavy-element masses of \citet{Chachan_2025}} for the 37 planets in both samples. Heavy-element masses are computed using $M_Z = Z_p M_p$ from their bulk metallicities $Z_p$.}
    \label{fig:chachan}
\end{figure}

Our core masses are benchmarked against the heavy-element masses of \cite{Chachan_2025} (hereafter C25), who infer bulk metallicities for a large sample of close-in giant planets ($20\,M_\oplus$ -- $20\,M_{\rm Jup}$). The C25 models place most metals uniformly in the gas-rich envelope, with up to $10\,M_\oplus$ of metals in the gas-free core. Using only planets with reported bulk metallicities $Z_p$ (excluding upper limits), we convert their $Z_p$ to metal masses using $M_Z = Z_p M_p$. Figure~\ref{fig:chachan} compares the $M_Z$ values inferred by C25 to the $M_{\rm core}$ values inferred by our model. Despite our planet model and that of C25 differing in their equations of state, atmospheric boundary conditions, and age priors, we find remarkable similarity in the $M_{\rm core}$ values from our work compared to the $M_Z$ values from C25: 65\% of the $M_{\rm core}-M_Z$ values agree within 1$\sigma$, and 95\% within 2$\sigma$. However, we do find our $M_{\rm core}$ estimates are systematically lower than the $M_Z$ values from C25 by $\sim$10--20\%. This could reasonably be attributed to differences in placement and densities of metals; all metals in an iron-silicon core in this work, versus a $\sim$10 M$_\oplus$ core with primarily water-ice metals spread through the atmosphere in C25.

Differences in our deduced planet ages from C25 can be attributed to our differences in priors: C25 places a gaussian prior using stellar ages when available, with uniform priors between 0.5--10 Gyr otherwise. A handful of planets have inferred ages younger than 0.5 Gyr, and are excluded from their sample. We instead adopt a uniform prior of 0.1--10 Gyr for all planets, and do not use stellar ages. We exclude any planets which require ages less than 0.1~Gyr to explain their radii: Appendix~\ref{sec:appendix_models} contains a list of such planets, as well as planets too dense to be described by an iron-silicon core (Tab.~\ref{tab:badfits}). Most inflated radii likely require anomalous power \citep[e.g.][]{ThorngrenFortney2018, Karalis+2025}, possibly due to Ohmic dissipation \citep[e.g.][]{PuValencia2017, Batygin2025} or residual heat from tidal circularization \citep[e.g.][]{YuDai2024, Hallatt+2026, Weldon+2026}. Combined, removing the fits listed in Appendix \ref{sec:appendix_models} in Table \ref{tab:badfits} revises our total sample size of Neptunes down from 231 to 215. 

\begin{figure}[t]
    \centering
\includegraphics[width=\columnwidth]{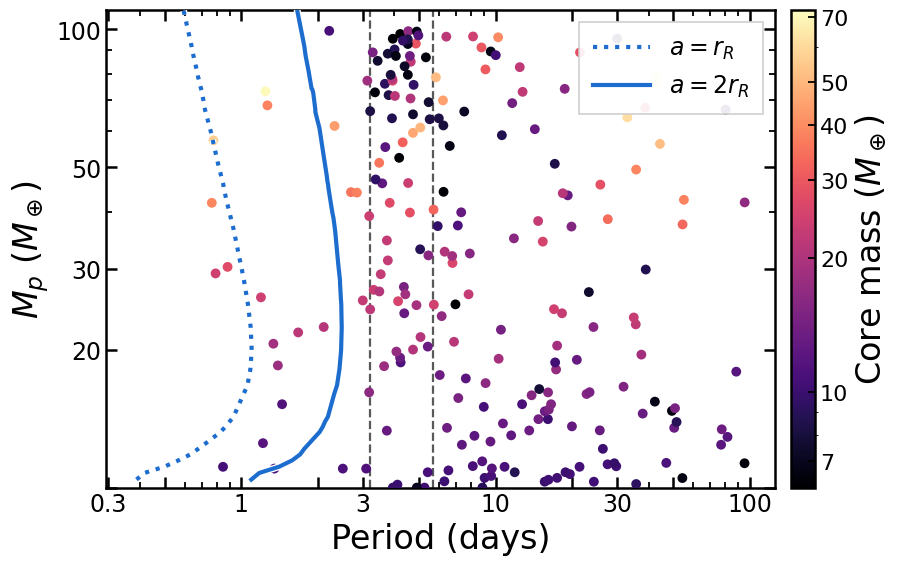}
    \caption{
    \textbf{Neptune core masses in $M_p$--$P_{\rm orb}$ space.} Planet mass versus orbital period, with points colored by the core mass from our MCMC fits to our planet structure model grid indicated in the color bar legend. Black dashed lines indicate the observational boundaries between Desert, Ridge, and Savanna Neptunes. Blue dotted and solid curves denote periods where the semi-major axis lies at one and two times the Roche radius $r_R$ for a solar host \citep{Eggleton1983}, computing these two radii for Neptunes with $10 M_\oplus$ cores at 5~Gyr ages.
    }
    \label{fig: M vs. P col core mass}
\end{figure}

We plot our $M_{\rm core}$ values in $M_p - P_{\rm orb}$ space in Figure~\ref{fig: M vs. P col core mass}. The orbital period corresponding to twice the Roche radius $r_R$ (for a solar host) is plotted in solid blue, illustrating that the location of the Neptune Ridge is consistent with the pile-up expected via high-eccentricity migration; most planets cluster outside the distribution limit $\sim r_R$, which circularize exterior to $\sim 2 r_R$. Also apparent are the large $M_{\rm core}$ values for Desert Neptunes, and the excess of several tens of $M_\oplus$ planets with $M_{\rm core}$ values clustered near $\sim 10 M_\oplus$ in the Savanna. The next section analyzes the statistical significance of these trends in more detail.

\section{Compositional Trends Across the Exo-Neptune Landscape}
\label{sec:results}

\begin{deluxetable*}{lccccccc}
\tablecaption{
Exoplanet sample analyzed in this work, alongside results from our model. Columns denote planet name, orbital period, planet radius, planet mass, core mass posteriors, host irradiation flux, and corrected Gaia DR3 metallicities (see Sec. \ref{sec:Methods} and~\ref{sec:results} for details). This table is available in its entirety in machine-readable form online, we provide an abridged version here for reference.
\label{tab:sample}}
\tablewidth{0pt}
\tabletypesize{\scriptsize}
\tablehead{
\colhead{Planet} &
\colhead{Period (d)} &
\colhead{Radius ($R_\oplus$)} &
\colhead{Mass ($M_\oplus$)} &
\colhead{Core Mass ($M_\oplus$)} &
\colhead{EMF} &
\colhead{$F_{\rm irrad}$ (erg s$^{-1}$ cm$^{-2}$)} &
\colhead{[M/H]}
}
\startdata
HATS-48 A b & $3.131667^{+0.000004}_{-0.000004}$ & $8.970^{+0.170}_{-0.170}$ & $77.20^{+7.00}_{-9.50}$ & $18.47^{+5.13}_{-3.82}$ & $0.76^{+0.05}_{-0.07}$ & $4.72^{+0.48}_{-0.48}\times10^{7}$ & \nodata \\
HATS-7 b & $3.185315^{+0.000005}_{-0.000005}$ & $6.310^{+0.520}_{-0.380}$ & $39.10^{+4.40}_{-4.40}$ & $23.83^{+4.33}_{-4.04}$ & $0.40^{+0.10}_{-0.11}$ & $7.63^{+0.76}_{-0.76}\times10^{7}$ & \nodata \\
Kepler-100 b & $1.6739035^{+0.0000004}_{-0.0000004}$ & $3.880^{+0.160}_{-0.160}$ & $21.80^{+6.20}_{-6.20}$ & $21.15^{+5.42}_{-5.45}$ & $0.10^{+0.25}_{-0.25}$ & $6.53^{+0.65}_{-0.68}\times10^{8}$ & $0.26^{+0.03}_{-0.03}$ \\
Kepler-411 b & $3.005156^{+0.000002}_{-0.000002}$ & $2.401^{+0.053}_{-0.053}$ & $25.60^{+2.60}_{-2.60}$ & $22.79^{+2.19}_{-2.39}$ & $0.01^{+0.09}_{-0.09}$ & $6.69^{+0.67}_{-0.67}\times10^{7}$ & $0.03^{+0.21}_{-0.06}$ \\
Kepler-94 b & $2.50806^{+0.00001}_{-0.00001}$ & $3.045^{+0.115}_{-0.115}$ & $11.00^{+2.60}_{-2.60}$ & $10.89^{+2.43}_{-2.29}$ & $0.05^{+0.21}_{-0.22}$ & $7.44^{+0.73}_{-0.73}\times10^{7}$ & $0.48^{+0.04}_{-0.07}$ \\
LTT 9779 b & $0.7920640^{+0.0000001}_{-0.0000001}$ & $4.720^{+0.230}_{-0.230}$ & $29.32^{+0.78}_{-0.81}$ & $24.16^{+0.98}_{-1.02}$ & $0.18^{+0.03}_{-0.03}$ & $8.78^{+0.87}_{-0.88}\times10^{8}$ & $0.34^{+0.01}_{-0.01}$ \\
NGTS-4 b & $1.337351^{+0.000008}_{-0.000008}$ & $3.180^{+0.950}_{-0.950}$ & $20.60^{+3.00}_{-3.00}$ & $18.82^{+3.08}_{-2.90}$ & $0.10^{+0.14}_{-0.15}$ & $3.24^{+0.33}_{-0.32}\times10^{8}$ & \nodata \\
TOI-1288 b & $2.699828^{+0.000005}_{-0.000005}$ & $4.960^{+0.100}_{-0.100}$ & $44.10^{+2.70}_{-2.60}$ & $35.28^{+2.47}_{-2.41}$ & $0.20^{+0.05}_{-0.06}$ & $1.70^{+0.17}_{-0.17}\times10^{8}$ & \nodata \\
TOI-132 b & $2.10970^{+0.00001}_{-0.00001}$ & $3.420^{+0.130}_{-0.140}$ & $22.40^{+1.90}_{-1.90}$ & $21.03^{+1.59}_{-1.93}$ & $0.07^{+0.09}_{-0.07}$ & $2.08^{+0.20}_{-0.20}\times10^{8}$ & $0.26^{+0.02}_{-0.01}$ \\
TOI-1410.01 & $1.21687^{+0.00003}_{-0.00003}$ & $3.100^{+0.220}_{-0.160}$ & $12.50^{+1.10}_{-1.00}$ & $11.95^{+1.03}_{-1.05}$ & $0.06^{+0.08}_{-0.08}$ & $1.92^{+0.19}_{-0.19}\times10^{8}$ & $0.11^{+0.01}_{-0.01}$ \\
\enddata
\end{deluxetable*}
\begin{figure*}[t]
\centering
\includegraphics[width=\textwidth]{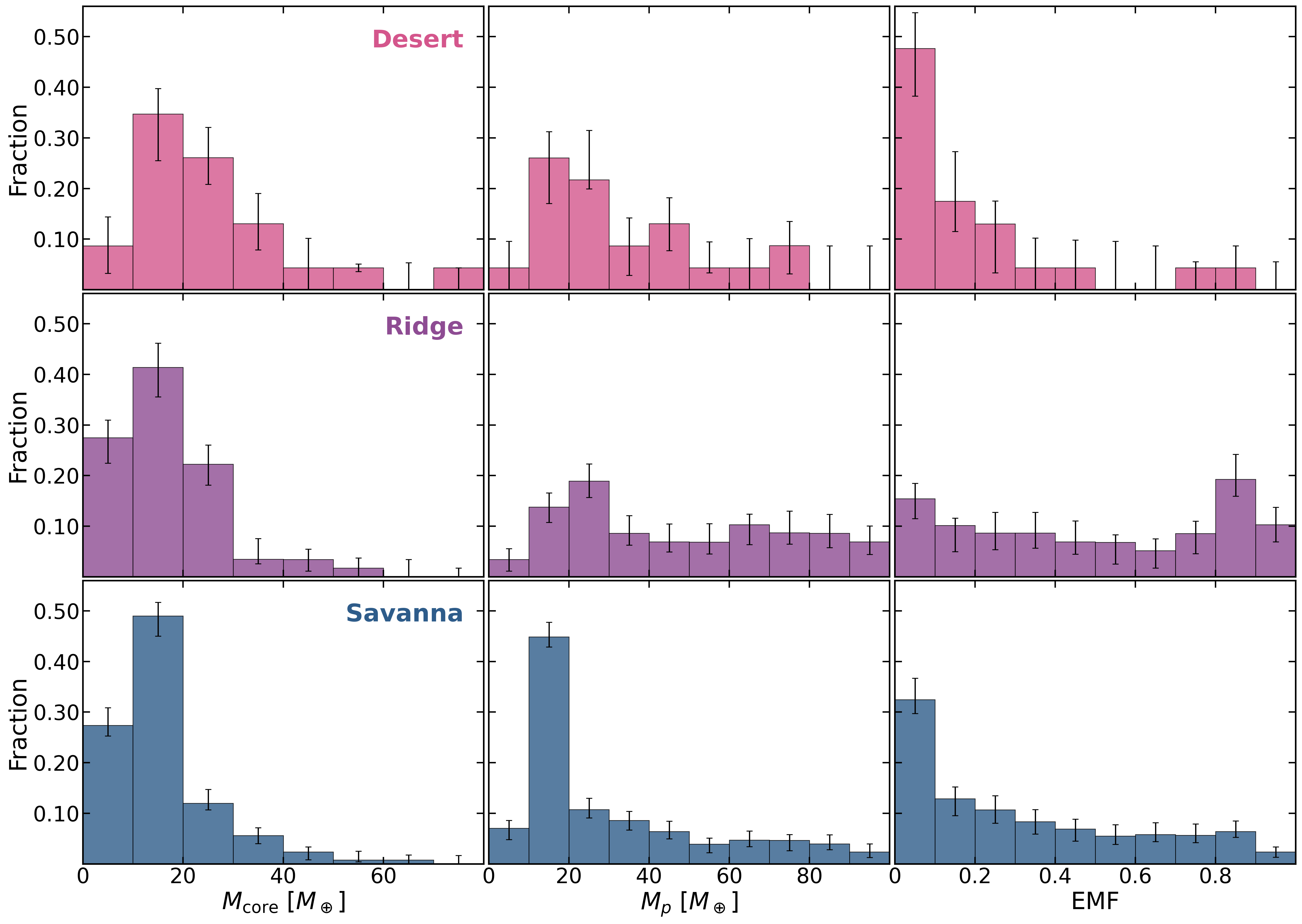}
    \caption{\textbf{Desert Neptunes have larger core masses, Ridge Neptunes have larger EMFs, and Savanna Neptunes have smaller planet masses.} Rows show the Desert, Ridge, and Savanna populations; columns show core mass $M_{\rm core}$, planet mass $M_p$, and Envelope Mass Fraction ${\rm EMF} = 1 - M_{\rm core}/M_p$. Bar heights and error bars are computed by Monte Carlo resampling of each planet's measurement uncertainty (see Sec. \ref{sec:results}). 
    }
    \label{fig: hist of neptunes}
\end{figure*}

\begin{figure*}[t]
    \centering
\includegraphics[width=\textwidth]{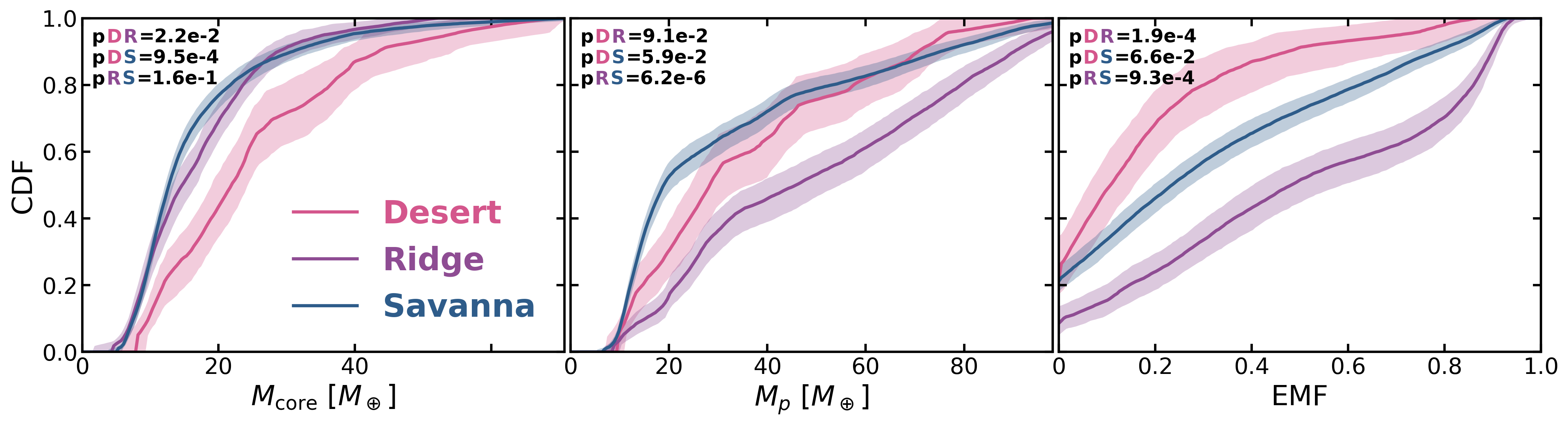}
    \caption{\textbf{Two-sided Kolmogorov-Smirnov (KS) tests confirm Desert Neptunes have larger core masses, Ridge Neptunes have larger EMFs, and Savanna Neptunes have smaller planet masses.} Left, middle, and right panels show the Cumulative Distribution Functions (CDFs) of the core mass, planet mass, and EMF, respectively, for the Desert (pink), Ridge (purple), and Savanna (blue) populations. For each of the 1000 iterations, we drew a bootstrap sample for each population, where we select planets at random with replacement, and also sample from the measurement or posterior uncertainties. Curves show the median CDF across iterations, and shaded bands show the 68\% credible interval at each point (Sec.~\ref{sec:results}). In the top left of each panel, $p_{\rm DR}$, $p_{\rm DS}$, and $p_{\rm RS}$ are the median two-sample KS test $p$-values across all iterations for the Desert--Ridge, Desert--Savanna, and Ridge--Savanna comparisons, respectively ($D, R, S =  {\rm Desert}, {\rm Ridge}, {\rm Savanna}$).}
    \label{fig: cdfs of neptuness}
\end{figure*}

Now we aim to quantify comparisons of core masses, planet masses, and EMFs for Desert, Ridge, and Savanna populations. We start by constructing histograms of these three planet parameters for each Neptune population in a Monte Carlo fashion (Fig.~\ref{fig: hist of neptunes}). For each planet, we draw 1000 values of $M_\mathrm{core}$ and $M_p$ from split-normal distributions set by the asymmetric uncertainties from our MCMC analysis, and compute the EMF of each draw as ${\rm EMF}= 1 - M_\mathrm{core}/M_p$. Combining the $i$th draw of every planet in a population gives 1000 realizations of each parameter histogram. Bar height in Figure~\ref{fig: hist of neptunes} is the mode of the distribution of heights, and error bars correspond to the 68\% highest posterior density interval (HPDI). We then calculate the cumulative distribution functions (CDFs) for each of the three planet parameters. These CDFs additionally include uncertainties due to sample size via bootstrapping. In short, for each of the 1000 iterations we resample with replacement; given a population of $N$ planet measurements, we draw $N$ measurements at random so that some planets appear more than once and others not at all for a given iteration. For each planet measurement drawn, we then include measurement uncertainties via the Monte Carlo method as described above. We compute the CDF of each iteration, and report the median curve and the 68\% credible interval of the 1000 CDFs at each value (Fig.~\ref{fig: cdfs of neptuness}).

\subsection{Desert Neptunes Have More Massive Cores}
\label{sucsec:massive cores}

Neptunes in the Desert have considerably larger cores than Neptunes in either the Ridge or Savanna. By eye, the left column of Figure~\ref{fig: hist of neptunes} shows the Ridge and Savanna core mass distributions more concentrated at 10--20$M_\oplus$ while the Desert distribution is broader; there is substantial weight out to and beyond $M_{\rm core}$ $\sim$ 30$M_\oplus$. Trends in median core mass further support larger cores for short-separation Neptunes; median $M_{\rm core}$ decreases from the Desert ($\sim$ 23$M_\oplus$) to the Ridge ($\sim$ 16$M_\oplus$) to the Savanna ($\sim$ 14$M_\oplus$). This higher median core mass of Desert Neptunes is not only driven by a larger probability density at higher masses, but also a lack of systems at the low-mass end of the distribution: none of the 23 Desert planets have a best-fit core mass below 10$M_\oplus$, compared to $\sim 27\%$ of Ridge planets and $\sim 19\%$  Savanna planets. The corresponding CDF, shown in the left panel of Figure~\ref{fig: cdfs of neptuness}, makes the offset of Desert Neptunes to higher core masses in comparison to the two other populations even more apparent. To quantify the significance of the differences between core mass distributions, we apply the two-sample Kolmogorov--Smirnov (KS) test from \texttt{SciPy} to each of the 1000 resampled pairs of populations, where each resample is drawn from the bootstrap and Monte Carlo procedure described in Section~\ref{sec:results}, and report the median $p$-value across all resamples. With this simple diagnostic we find that the Desert and Savanna core mass distributions are significantly different, with a median $p$-value of $9.5 \times 10^{-4}$ and 91\% of the resamples yielding $p < 0.05$ ($\sim$2$\sigma$ significance). The Desert and the Ridge distributions are also significantly different (median $p$-value of 0.022), though less decisively so, with 62\% of the resamples reaching $p < 0.05$. By contrast, the Ridge and Savanna $M_{\rm core}$ CDFs are statistically indistinguishable (median $p$-value of 0.16; 28\% of resamples yield $p < 0.05$).

\subsection{Ridge Neptunes Have More Massive Envelopes}

Ridge Neptunes display disproportionate EMFs relative to Savanna and Desert Neptunes.
In the rightmost column of Figure \ref{fig: hist of neptunes}, both the Desert and Savanna distributions have significant peaks near EMF $\sim$ 0, populated largely by rocky planets with low-mass atmospheres. In contrast, the Ridge EMF histogram is largely flat. The stark differences in the EMF distributions for Neptunes in the Ridge, relative to those in the Desert and Savanna, are most apparent when examining the CDFs (Fig.~\ref{fig: cdfs of neptuness} right panel). The Ridge CDF is shifted toward substantially higher EMFs with little curvature relative to the Desert and Savanna, reflecting a flat EMF distribution. Two sample KS tests yield median $p$-values of $1.9 \times 10^{-4}$ and $9.3 \times 10^{-4}$, with 98$\%$ and 92$\%$ of resampled CDF pairs achieving $p<0.05$, when comparing the Ridge to the Desert and Savanna CDFs, respectively. This is strong evidence that the Ridge EMF distribution is distinct from both the Desert and Savanna populations. The Desert and Savanna have modestly similar EMF distributions, with the KS test returning a median $p$-value of 0.066, with only 45\% of resamples reaching $>2\sigma$ significance when comparing these two population CDFs.

\subsection{Savanna Neptunes Have Smaller Planet Masses}
\label{subsec: savanna stuff}

Neptunes in the Savanna are systematically less massive than those in the Desert and Ridge. The middle column of Figure \ref{fig: hist of neptunes} shows Savanna planets with a substantial pile-up in the 10--$20\,M_\oplus$ bin, whereas the Desert and Ridge masses are more distributed. This low-mass peak for Savanna planets is clearly evidenced in the middle panel of Figure \ref{fig: cdfs of neptuness}, where the Savanna CDF climbs most steeply and sits to the left of the other two CDFs before 20$M_\oplus$. A two-sample KS test for the Savanna--Ridge pair yield a median $p$-value of $6.2\times10^{-6}$, with 100\% of resamples yielding $p < 0.05$, while the Desert--Savanna pair comparison exhibits more modest distinction with median $p$-value of $5.9\times10^{-2}$ and 47\% of resamples producing $p < 0.05$.

\subsection{Savanna Neptune Compositions Hinge on Host Star Metallicity}

\begin{figure}[t]
    \centering
\includegraphics[width=\columnwidth]{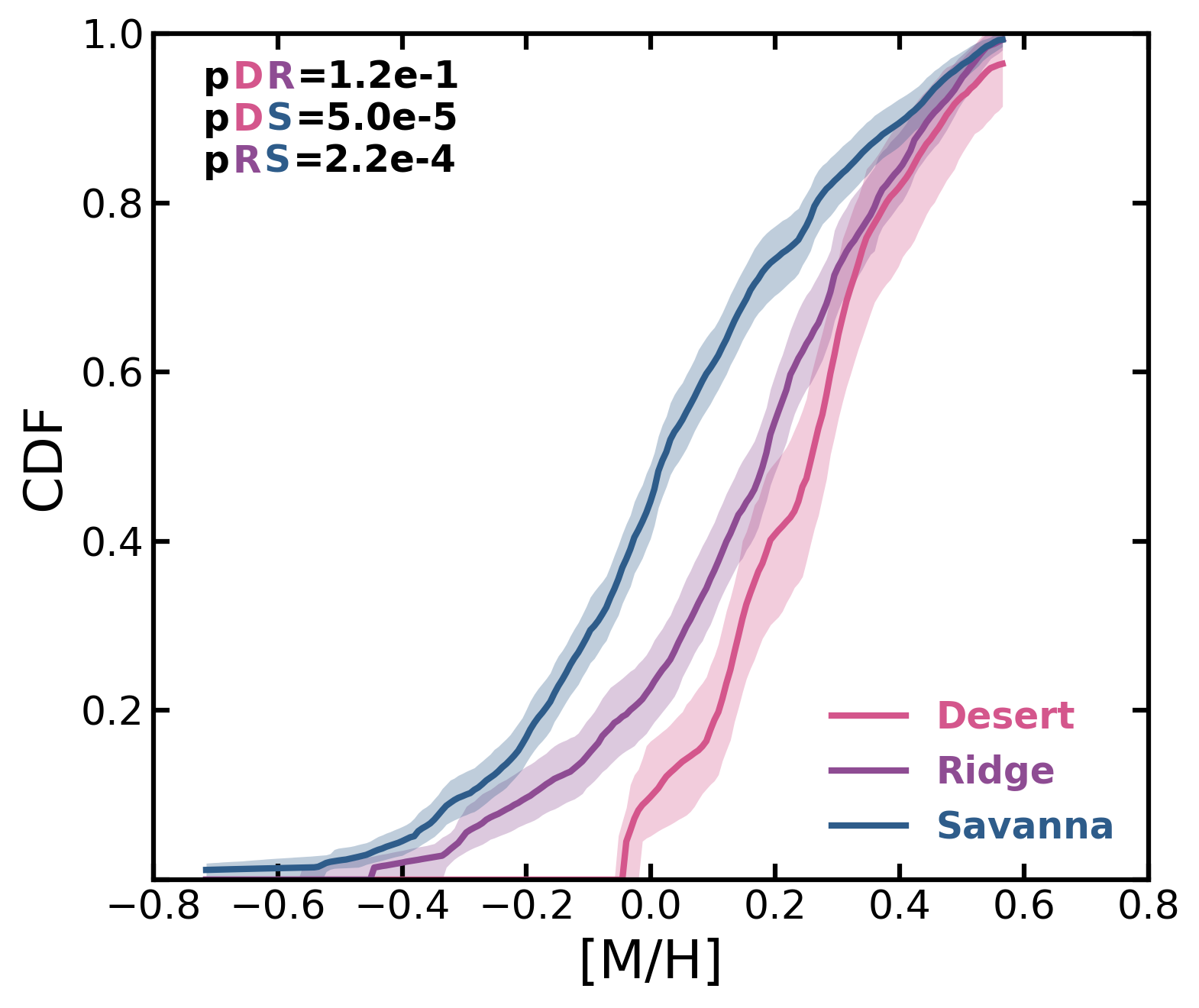}
    \caption{\textbf{Desert and Ridge Neptunes orbit more metal-rich host stars than Savanna Neptunes.} Displayed are CDFs of host star Gaia DR3 metallicities [M/H] for Desert, Ridge, and Savanna Neptunes. Top left displays median two-sample KS $p$-values computed between the three population CDFs. The confidence interval and $p$-value calculations, as well as Neptune population label colors and notations, are identical to those in Fig.~\ref{fig: cdfs of neptuness}.}
    \label{fig: met cef}
\end{figure}

\begin{figure*}[t]
    \centering
\includegraphics[width=\textwidth]{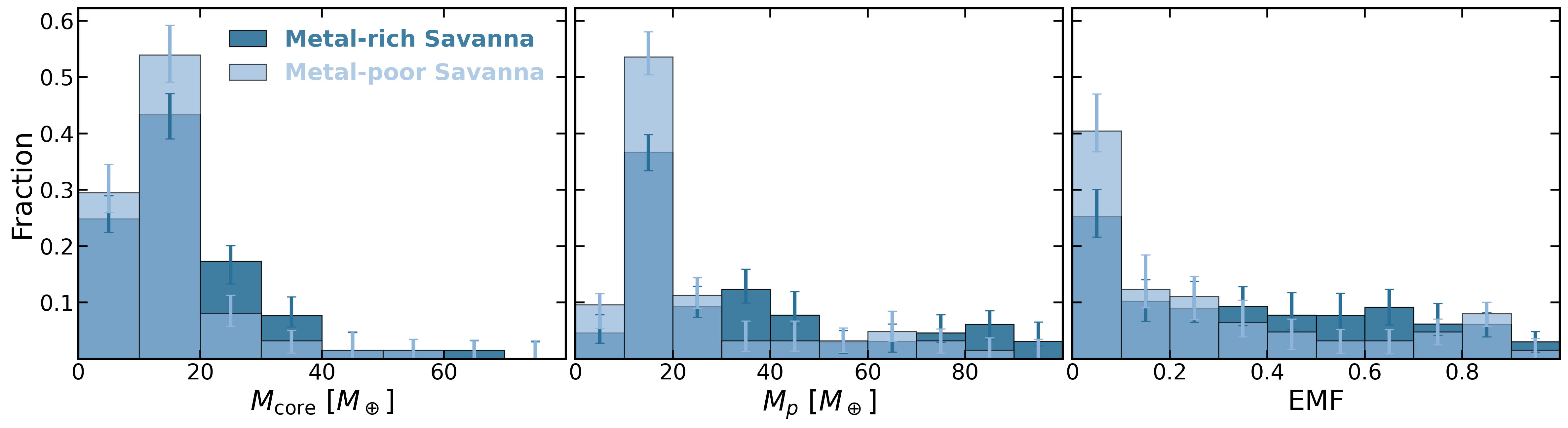}
    \caption{\textbf{Metal-poor Savanna hosts favor lower core masses, lower planet masses, and lower EMFs than metal-rich Savanna hosts.} Left, middle, and right panels show core mass, planet mass, and EMF, respectively, for Savanna Neptunes with hosts that are metal-poor ([M/H] $\leq0$) and metal-rich ([M/H] $>0$). Bar heights are the mode, and error bars denote the 68\% HPDI (see Sec.\ref{sec:results} for details). }
    \label{fig: hist of sav2}
\end{figure*}

We query the Gaia DR3 catalogue to obtain homogeneous metallicity measurements of the host stars in our sample of Neptunes, exclusively selecting those derived from the spectroscopic pipeline (GSP--Spec), available for 19 Desert, 56 Ridge, and 106 Savanna Neptunes. We apply the \cite{YeeWinn2023} calibration to Gaia DR3 GSP--Spec values $[{\rm M/H}]_{\rm GSP-Spec}$ to obtain host metallicities $[{\rm M/H}]$,
\begin{align}
    &[{\rm M/H}] = 0.384 + 0.761 \cdot [{\rm M/H}]_{\rm GSP-Spec} \nonumber \\
    &- 0.392 \cdot ({\rm BP} - {\rm RP}) 
    \nonumber \\
    &+ 0.708 \cdot  ({\rm BP} - {\rm RP} - 0.95) \cdot H({\rm BP} - {\rm RP} - 0.95),
\end{align}
where ${\rm BP} - {\rm RP}$ is the Gaia $G_{\rm BP} - G_{\rm RP}$ color in magnitude, with $H(x)$ the Heavyside step function. We list $[{\rm M/H}]$ values in Table~\ref{tab:sample}.

Host star metallicity separates the Savanna from the Ridge and the Desert populations (Fig.~\ref{fig: met cef}). While Ridge metallicities start at lower values than the Desert, with 8 of the 56 Ridge hosts (14\%) having sub-solar metallicities, both populations have super-solar median values, [M/H] = 0.26 and [M/H] = 0.19 for the Desert and Ridge host stars respectively. Conversely, the Savanna host star metallicity distribution is more evenly distributed about solar, with a median [M/H] = 0.03. A KS test quantifies this; while the Desert and Ridge are statistically indistinguishable (median $p = 0.12$, with 32\% of resamples yielding $p<0.05$), the Savanna diverges significantly from both the Desert (median $p = 5.0\times10^{-5}$, with 100\% of resamples yielding $p<0.05$) and the Ridge (median $p = 2.2\times10^{-4}$, with 96\% of resamples yielding $p<0.05$).

While nearly all Desert and Ridge hosts are metal-rich, the Savanna has a largely even distribution on either side of solar. How might this broader host star metallicity distribution imprint on the composition of the Savanna planets? We divide the Savanna sample into metal-rich ($[{\rm M/H}] > 0$; 63 planets) and metal poor ($[{\rm M/H}] \leq 0$; 43 planets) and ask whether core mass, planet mass, and EMF distributions differ as a function of host star metallicity. We find that while distribution shapes are similar,  metal-poor Savanna systems host Neptunes with systematically smaller core masses, planet masses, and EMFs (Fig.~\ref{fig: hist of sav2}).

\section{Distinct Origins of the Desert, Ridge, and Savanna} 
\label{sec:disc}

The Desert, Ridge, and Savanna planets exhibit stark differences in their core masses, planet masses, and EMFs. We find that Desert Neptunes have higher core masses than Ridge or Savanna planets along with an EMF distribution sharply peaked at small values. While Ridge Neptunes have similar total masses as their Desert counterparts, much of this mass is in their atmospheres rather than their cores. Savanna planets stand out with smaller planet masses than either the Ridge or Desert Neptunes, with an EMF distribution shaped by host star metallicity.

Elsewhere, growing observational evidence also supports Desert and Ridge Neptunes sharing dynamically boisterous pasts with hot Jupiters, while the pasts of Savanna Neptunes might be more quiescent. Observations show metal-rich stars host systems with more dynamically-hot architectures, sporting larger eccentricities \citep[e.g.][]{Morgan+2026} and mutual inclinations \citep[e.g.][]{Hua+2025}. As discussed in \cite{Vissapragada_2025}, the host metallicities of hot Jupiters, Ridge Neptunes, and Desert Neptunes are all similarly metal-rich, with metallicities significantly higher than Savanna Neptunes, short-separation sub-Neptunes/super-Earths, and longer-period small planets. Stellar obliquities of Ridge Neptunes have a similar distribution as those of hot Jupiters \citep[e.g.][]{Espinoza-Retamal+2026}, and a lone Desert obliquity measurement revealed a Neptune on a polar orbit \citep{Bourrier2022}. Similar to hot Jupiters, which lack the nearby close companions of warm Jupiter systems \citep[e.g.][]{Huang+2016, Wu+2023}, the close companion frequency of Desert and Ridge Neptunes is notably smaller than that of Savanna Neptunes \citep{Thomas_2026}, implying hot Jupiters, Desert Neptunes, and Ridge Neptunes could have dynamically emptied their environments. Furthermore, a substantial, $\sim$30--50\% subset of Desert Neptunes have eccentricities larger than $e\gtrsim 0.1-0.2$ \citep[e.g.][]{Correia_2020, PetrovichBhaskar2026}.

In what follows, we discuss the processes that could produce Neptunes in the Desert, Ridge, and Savanna, with the compositions and dynamical histories implied by their bulk densities, orbital architectures, and host star metallicities.

\subsection{Neptune Savanna: Gas-Poor Formation In-Situ}

\begin{figure}[t]
    \centering
\includegraphics[width=\columnwidth]{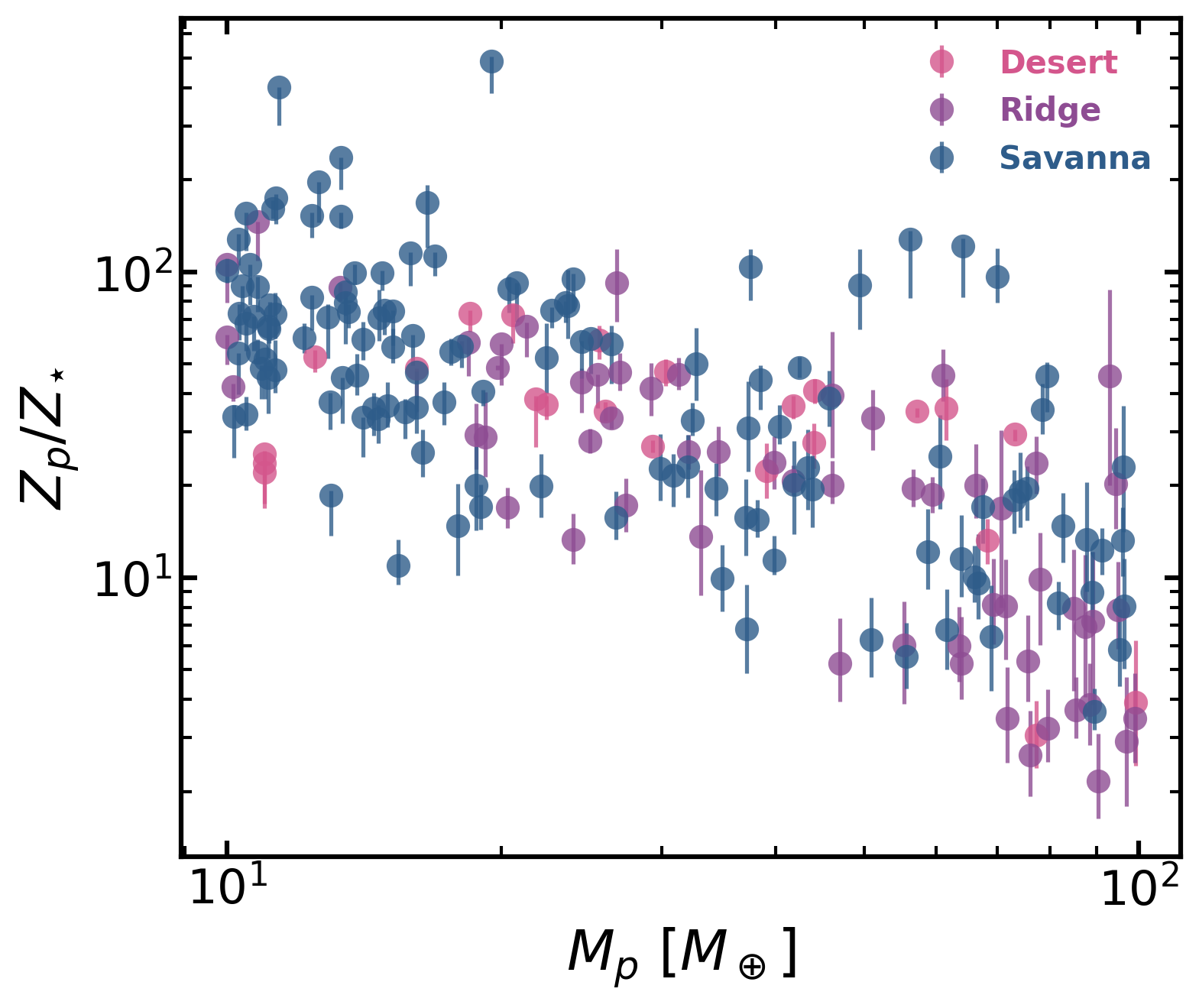}
    \caption{\textbf{The Neptune mass-metallicity relation.} Planet bulk metallicity $Z_p \simeq M_{\rm core}/M_p$ over the stellar metallicity $Z_\star$ plotted against planet mass, for the Desert, Ridge, and Savanna. We compute the stellar metal mass fraction using $Z_\star = Z_\odot\,10^{[{\rm M/H}]}$ with $Z_\star = 0.014$ \citep{Asplund+2009}. Error bars are the median and 68\% credible interval from 1000 Monte Carlo realizations of $M_{\rm core}$, $M_p$, $Z_\star$, and $Z_p$, drawn from asymmetric uncertainties.}
    \label{fig: metal vs. mass}
\end{figure}

We argue the Savanna population provides an important baseline of Neptunes that formed primarily \textit{in-situ}. A strong peak in the planet mass distribution near 10--15 M$_\oplus$ reflects the rough maximum expected core size before runaway gas accretion forms a more massive planet \citep[e.g.][]{Pollack1996, Batygin+2016}. Furthermore, the overall distribution at sub-100-day periods reflects the expected picture of sub-Saturn growth in gas-poor environments; scant solid material only builds $\gtrsim 10 M_\oplus$ cores late, with little time to grow EMFs larger than $\gtrsim10\%$ \citep[e.g.][]{Lee_2019}. As gas depletes, while cores migrate little via dynamical friction, orbital instability could lead to core growth $\gtrsim 10-20 M_\oplus$ masses through major mergers \citep[e.g.][]{LeeChiang2016}. The broad range of Savanna host metallicities, causing large scatter in $Z_p/Z_\star$ (Figure~\ref{fig: metal vs. mass}) is consistent with the independence of small-planet occurrence from metallicity \citep{Petigura+2018}. In particular, metal-poor Savanna hosts have smaller cores, lower total masses, and lower EMFs than metal-rich Savanna hosts (see Fig.~\ref{fig: hist of sav2}), whose effect on solid content and nebulae metallically could lead to longer growth timescales, smaller core masses, and shorter gas accretion phases resulting in overall gas-poor/low EMF planets \citep[e.g.][]{LeeChiang2015, Lee_2019, Chachan+2021}

\subsection{Neptune Ridge: Gas-Rich Formation Followed by High-Eccentricity Migration}

Moving inwards, the Ridge planets have core masses similar to those of the Savanna, but have substantially more mass in their gaseous envelopes and preferentially orbit metal-rich host stars when compared to Savanna hosts. Combined, these results point to formation farther from their host stars followed by high-eccentricity migration, which placed these Neptunes at their current locations. Forming farther out provides a larger reservoir of solids to quickly grow cores and accrete the substantial gaseous envelopes seen in Ridge Neptunes \citep[e.g.][]{Lee_2019, Chachan+2021}, which contrast strongly with the small, $\lesssim 10\%$ EMFs of Savanna and Desert Neptunes. Subsequent migration of Ridge Neptunes via dynamical interactions with a companion planet or star, excitation onto a high-eccentricity orbit, and finally tidal circularization could place these Ridge Neptunes on their present day orbits. Circularization via dynamical tides following surface gravity wave shocks is a particularly appealing way to not only explain the orbital clustering of the Neptune Ridge, but also the hot Jupiter pile-up \citep{Zanazzi_2026}.

We find other mechanisms to deliver Neptunes to the Ridge less viable. Radial migration to the truncated protostellar inner disk edge might deliver Neptunes to these short separations \citep[e.g.][]{BaileyBatygin2018, Batygin+2023}, but does not explain the enhanced metallicities of Ridge Neptune hosts relative to hosts of longer separation, or less massive, planets (\citealt{Vissapragada_2025}, see also Fig.~\ref{fig: met cef}). The notion that Ridge Neptunes might have first been hot Jupiters, but were down-sized after their envelopes were tidally-stripped by the host star \citep{Weldon+2026}, lies in tension with their typical core masses of $\sim10-20 M_\oplus$, much smaller than the metal masses of most gas giants, but roughly consistent with the gaseous planet mass-metallicity relation (Fig.~\ref{fig: metal vs. mass}, e.g.~\citealt{Thorngren_2016, Chachan_2025})

\subsection{Neptune Desert: Barren Jovian Cores}

Within the Neptune Desert, we find planets have the largest core masses of the three populations as well as the smallest EMFs, peaking even more strongly at low EMFs than the Savanna population. In addition, both \citet{Vissapragada_2025} and our work find that Desert Neptunes have super-solar metallicities consistent with the Ridge population and substantially above the Savanna planets (Fig.~\ref{fig: met cef}). \citet{Vissapragada_2025} additionally finds consistent metallicity distributions between Desert Neptune and hot Jupiter host stars. 

Even with the large host star metallicity, a possible sign of an increased solids budget (Fig.~\ref{fig: metal vs. mass}), growing $\gtrsim20-50 M_\oplus$ cores with EMFs below $\lesssim 10\%$ is a colossal feat for core-accretion alone to accomplish, since core masses significantly above $\sim 10 M_\oplus$ trigger run-away gas accretion \citep[e.g.][]{Batygin+2016, Lee_2019}. Instead, the large cores, small EMFs, and metal-rich hosts can all be explained if Desert Neptunes are the barren cores of former hot-Jupiters \citep{Vissapragada_2025}. In particular, the same process that can cluster Ridge Neptunes and pile-up hot Jupiters can also unbind gas-giant envelopes; when circularizing via dynamical tides, sufficiently eccentric orbits excite surface gravity waves that shock deep, which drive a hydrodynamic outflow that leaves barren, Jovian cores in the Neptune Desert \citep{Zanazzi_2026}. Alternatively, hot Jupiter cores might be placed in the Desert at very late ages, once hot Jupiters in-spiral due to tides raised in the host star, placing them on orbits susceptible to tidal stripping \citep[e.g.][]{Valsecchi_2014, Thorngren+2023, Hallatt_2026}. While the trajectory of hot-Jupiter-to-Desert-Neptune remains opaque, we note the ability of HEM followed by dynamical tides to explain the hot Jupiter pile-up, Neptune Ridge, and Neptune Desert within the same framework is compelling. Future work to test Desert Neptune origins are briefly discussed in the final section.

\section{Summary and Outlook}
\label{sec:conc}

In this work we assembled a sample of 215 Neptunes (10--100 $M_\oplus$) with measured masses and radii across the Neptune Desert ($P < 3.2$ d), Ridge ($3.2 < P < 5.7$ d), and Savanna ($5.7 < P < 100$ d) to quantify bulk compositions of each planet, identify population-level trends, and connect these to similar or distinct formation and evolutionary pathways. After constructing a grid of  planetary evolutionary models, we implement a MCMC framework to compute posterior distributions for core masses and EMFs, given the observed planetary masses and radii. 

Our histograms (Fig.~\ref{fig: hist of neptunes}) and Cumulative Distribution Functions (CDFs; Fig.~\ref{fig: cdfs of neptuness}) that incorporate measurement and finite sample-size uncertainties compare core masses, EMFs, and planet masses across the three populations. We find that each Neptune population has an emblematic compositional trait: 1) Desert Neptunes have more massive cores; 2) Ridge Neptunes have more massive envelopes; and 3) Savanna Neptunes have smaller planet masses. In addition, we queried the Gaia DR3 table for GSP-Spec metallicities to compare the host star metallicity distributions of the three Neptune populations. We reaffirm prior results finding super-solar metallicities for Desert and Ridge Neptunes, with Savanna Neptune hosts centered instead around solar values. Savanna Neptune compositions are shaped by the broader distribution of stellar metallicities; metal-poor Savanna Neptunes have smaller cores, lower planet masses, and smaller EMFs than their metal-rich counterparts. 

We assert that our compositional trends combined with complementary lines of observational evidence point towards distinct origins of the Desert, Ridge, and Savanna. The Neptune Savanna is our baseline population, forming \textit{in-situ} in a gas-poor environment. Ridge planets likely formed farther out in a gas-rich environment, and underwent high-eccentricity migration to land at their present-day orbits near $\sim$2 times the Roche radius. Finally, the Neptune Desert is populated with stripped cores of former gas giants. While there are two primary explanations for placing these barren Jovian cores in the Desert, namely high-eccentricity migration followed by dynamical tides (HEM; \citealt{Zanazzi_2026}) and Roche-lobe overflow (RLO; \citealt{Hallatt_2026}), we find the ability of HEM to also explain the hot Jupiter pile-up and Neptune Ridge compelling.

The pathway Desert Neptunes take from gas giant to stripped core remains a key open question. Timely endeavors to crack the origins of Desert planets include stellar obliquity measurements of a population of Desert Neptunes via the Rossiter-McLaughlin effect; high-eccentricity migration (HEM) predicts misaligned obliquities \citep{Zanazzi_2026}, while Roche-Lobe overflow (RLO) predicts aligned obliquities \citep{Hallatt+2026_obliquity}. While there is one measured Desert Neptune obliquity showing a planet on a polar orbit \citep{Bourrier2022}, more are needed. Diverging predictions of outer giant companion fractions for HEM vs RLO additionally presents a clear observational test. A direct extension of this present work to the Desert Neptune origins question is to apply our MESA planetary evolution models to non-inflated hot and warm Jupiters, comparing our characterizations of bulk composition amongst the Neptune and Jupiter populations in this consistent framework. We will tackle this next step in a future paper.

\section{Acknowledgements}

We thank Yayaati Chachan and Shreyas Vissapragada for useful conversations. MLB acknowledges support from NSERC. JJZ thanks the Heising-Simons Foundation for support through an Early-Career Grant. The Center for Exoplanets and Habitable Worlds is supported by the Pennsylvania State University and the Eberly College of Science.

\facilities{NASA Exoplanet Archive \citep{NASAArchive}}

\software{\texttt{emcee} \citep{emcee}, \texttt{MESA} \citep{Paxton_2011, Paxton_2013, Paxton_2015, Paxton_2018, Paxton_2019}, \texttt{MESPA} \citep{MESPA}, \texttt{numpy} \citep{numpy}, \texttt{pandas} \citep{pandas}, \texttt{SciPy} \citep{scipy}, \texttt{matplotlib} \citep{Hunter_2007}}

\appendix
\restartappendixnumbering
\section{Population Dependence on Stellar Mass}
\label{sec:appendix_stellarmass}

\begin{figure}[t]
    \centering
\includegraphics[width=\columnwidth]{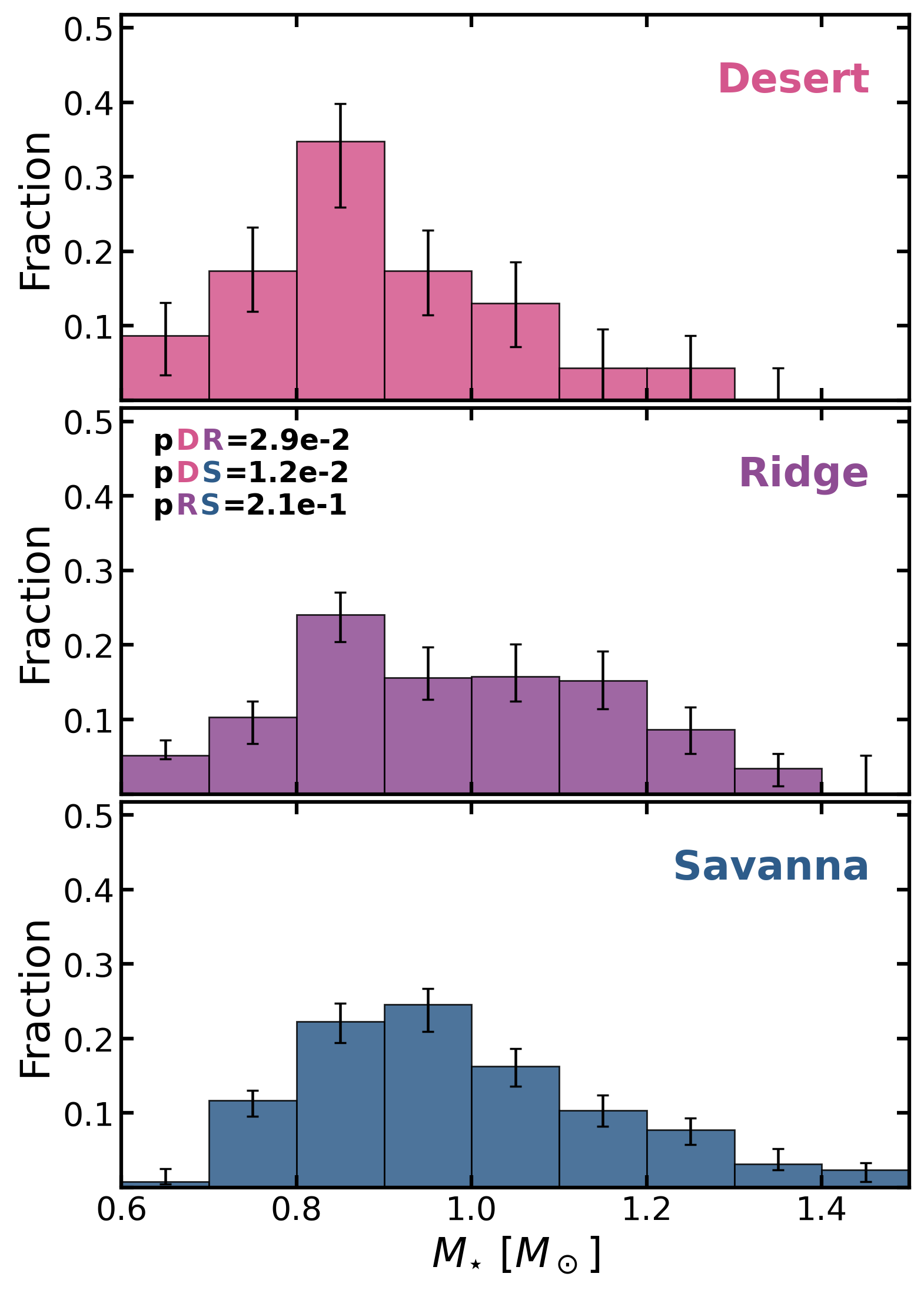}
    \caption{\textbf{The Ridge and Savanna have similar host star masses, both slightly higher than the Desert.} Bar heights are the mode and error bars the 68\% HPDI. Our sample contains no Neptune-mass planets with host star masses below $0.6\,M_\odot$.}
    \label{fig: host star mass dist}
\end{figure}

There is growing interest in the possible dependence of the three Neptune populations on the mass of the host star \citep[e.g.][]{Cui+2026_ridgehost}. Figure~\ref{fig: host star mass dist} displays the probability distributions for Neptunes in the Desert, Ridge, and Savanna, with the mass of the host star. Although all three seem roughly similar, we find slightly more Desert Neptunes around low-mass hosts. Median host star mass is similar for the Ridge ($0.959\,M_\odot$) and Savanna ($0.958\,M_\odot$), and lower for the Desert ($0.868\,M_\odot$). We note that we explicitly exclude M-dwarfs in our sample, but \cite{Thomas_2026} find that Desert and Ridge hosts extend into the M-dwarf regime. The few Neptunes detected around M-dwarfs are unlikely to alter our main results.

\section{Model parameters of failed MESA models}
\label{sec:appendix_models}

\begin{deluxetable}{cl}
\tablecaption{Grid models that failed to converge, listed by EMF and core mass.
\label{tab:missing_models}}
\tablewidth{0pt}
\tabletypesize{\scriptsize}
\tablehead{\colhead{EMF (\%)} & \colhead{$M_{\rm core}$ ($M_\oplus$)}}
\startdata
0.5 & 10$^*$, 15, 20, 25, 30, 35, 40, 45$^*$, 50$^*$, 55, 60, 65, 70, 75, 80, 85, 90, 95 \\
1   & 10, 15, 20, 25, 30, 35, 65 \\
20  & 10$^*$ \\
35  & 10$^*$, 30$^*$ \\
40  & 6, 8$^*$, 9$^*$ \\
45  & 6 \\
50  & 8$^*$ \\
60  & 8$^*$ \\
\enddata
\tablecomments{$^*$ denotes that irradiation failed for this core mass}
\end{deluxetable}

\begin{deluxetable*}{lccccccc}
\tablecaption{Planets with radii either too small or too large to be fit by our models. Columns display planet name, which population the planet is a part of, measured radii, posteriors of our attempted radii fits, the magnitude of the mismatch, also with the posterior medians for the core mass, envelope mass fraction, and age implied by our models.
\label{tab:badfits}}
\tablewidth{0pt}
\tabletypesize{\scriptsize}
\tablehead{
\colhead{Planet} &
\colhead{Type} &
\colhead{ Planet Radius ($R_\oplus$)} &
\colhead{Fitted Radius ($R_\oplus$)} &
\colhead{$(R_{\rm fit}-R_{\rm obs})/\sigma$} &
\colhead{Core Mass ($M_\oplus$)} &
\colhead{EMF} &
\colhead{Age (Gyr)}
}
\startdata
Kepler-107 c & Ridge & $1.597^{+0.026}_{-0.026}$ & $2.243$ & $+24.8$ & $7.11$ & $0.01$ & $6.42$ \\
WASP-127 b & Ridge & $14.690^{+0.300}_{-0.300}$ & $11.650$ & $-10.1$ & $6.09$ & $0.89$ & $0.11$ \\
HAT-P-51 b & Ridge & $13.510^{+0.190}_{-0.190}$ & $11.853$ & $-8.7$ & $6.13$ & $0.94$ & $0.11$ \\
TOI-1296 b & Ridge & $14.120^{+0.300}_{-0.300}$ & $11.877$ & $-7.5$ & $6.18$ & $0.94$ & $0.12$ \\
WASP-193 b & Savanna & $16.410^{+0.661}_{-0.661}$ & $11.509$ & $-7.4$ & $6.23$ & $0.87$ & $0.14$ \\
WASP-20 b & Ridge & $16.390^{+0.660}_{-0.660}$ & $11.717$ & $-7.1$ & $6.68$ & $0.93$ & $0.15$ \\
TOI-1347 b & Desert & $1.800^{+0.100}_{-0.100}$ & $2.415$ & $+6.1$ & $10.63$ & $0.01$ & $5.28$ \\
WASP-39 b & Ridge & $14.240^{+0.450}_{-0.450}$ & $11.672$ & $-5.7$ & $6.63$ & $0.93$ & $0.14$ \\
TOI-2583 A b & Ridge & $14.500^{+0.500}_{-0.500}$ & $11.728$ & $-5.5$ & $6.63$ & $0.93$ & $0.14$ \\
WASP-131 b & Ridge & $13.740^{+0.410}_{-0.410}$ & $11.705$ & $-5.0$ & $6.46$ & $0.93$ & $0.14$ \\
NGTS-26 b & Ridge & $14.890^{+0.660}_{-0.660}$ & $11.666$ & $-4.9$ & $7.04$ & $0.93$ & $0.17$ \\
NGTS-5 b & Ridge & $12.730^{+0.260}_{-0.260}$ & $11.620$ & $-4.3$ & $6.41$ & $0.92$ & $0.13$ \\
KELT-11 b & Ridge & $15.100^{+1.100}_{-1.100}$ & $11.339$ & $-3.4$ & $7.51$ & $0.88$ & $0.33$ \\
WASP-84 c & Desert & $1.950^{+0.120}_{-0.120}$ & $2.335$ & $+3.2$ & $11.40$ & $0.01$ & $5.42$ \\
HATS-43 b & Ridge & $13.230^{+0.560}_{-0.560}$ & $11.490$ & $-3.1$ & $7.30$ & $0.92$ & $0.19$ \\
TOI-1420 b & Savanna & $11.890^{+0.330}_{-0.330}$ & $10.887$ & $-3.0$ & $6.20$ & $0.79$ & $0.14$ \\
\enddata
\end{deluxetable*}

In this work, we sometimes find that our planetary structure models fail for two different reasons: (1) MESA fails to converge on a model in hydrostatic equilibrium for select values of the EMF and $M_{\rm core}$, and (2) our models fail to reproduce the radii, given the masses of an exoplanet subset. Table~\ref{tab:missing_models} lists EMF and $M_{\rm core}$ values for failure (1): with the exception of EMF values less than $\lesssim 1\%$, most grid values successfully converge on a planet in hydrostatic equilibrium (compare to Tab.~\ref{tab:grid}). Table~\ref{tab:badfits} lists the observed planets for which failure (2) occurs. Most models that fail have large radii, and are located in the Neptune Ridge. Anomalous power due to some unincluded heat source might be able to reconcile model radii with those observed \citep[e.g.][]{ThorngrenFortney2018, Karalis+2025}.

\bibliography{sample701}{}
\bibliographystyle{aasjournalv7}
\end{document}